\documentclass[aps,prx,twocolumn,superscriptaddress,showpacs]{revtex4-1}
\usepackage{amsmath,amssymb}
\usepackage{graphicx}
\usepackage{natbib}
\usepackage{color}

\begin{document}

\title{Coherent perfect absorbers: linear control of light with light}

\author{Denis~G.~Baranov}
\email[]{denisb@chalmers.se}
\affiliation{Department of Physics, Chalmers University of Technology, 412 96 Gothenburg, Sweden}
\affiliation{Moscow Institute of Physics and Technology, Dolgoprudny 141700, Russia}
\affiliation{ITMO University, St.~Petersburg 197101, Russia}

\author{Alex~Krasnok}
\affiliation{Department of Electrical and Computer Engineering, The University of Texas at Austin, Austin, Texas 78712, USA}

\author{Timur~Shegai}
\affiliation{Department of Physics, Chalmers University of Technology, 412 96 Gothenburg, Sweden}

\author{Andrea~Al\`u}
\affiliation{Department of Electrical and Computer Engineering, The University of Texas at Austin, Austin, Texas 78712, USA}

\author{Y.~D.~Chong}
\email[]{yidong@ntu.edu.sg}
\affiliation{Centre for Disruptive Photonic Technologies and
School of Physical and Mathematical Sciences, Nanyang Technological University, Singapore 637371, Singapore}

\begin{abstract}
Absorption of electromagnetic energy by a material is a phenomenon that underlies many applied problems, including molecular sensing, photocurrent generation and photodetection. Commonly, the incident energy is delivered to the system through a single channel, for example by a plane wave incident on one side of an absorber. However, absorption can be made much more efficient by exploiting wave interference. A coherent perfect absorber is a system in which complete absorption of electromagnetic radiation is achieved by controlling the interference of multiple incident waves. Here, we review recent advances in the design and applications of such devices. We present the theoretical principles underlying the phenomenon of coherent perfect absorption and give an overview of the photonic structures in which it can be realized, including planar and guided-mode structures, graphene-based systems, parity- and time-symmetric structures, 3D structures and quantum-mechanical systems. We then discuss possible applications of coherent perfect absorption in nanophotonics and, finally, we survey the perspectives for the future of this field.
\end{abstract}

\maketitle


\section*{Introduction}

Interference is a ubiquitous wave phenomenon occurring whenever two or more coherent waves overlap, leading to a spatial redistribution of energy. It underlies the operation of numerous devices, including antennas, interferometers, spectrometers and Bragg reflectors~\cite{SalehTeich, Haus, Ishimaru}.  One interesting recent discovery in optics and photonics concerns the role that interference can have in regulating light absorption. In particular, it is possible for a device to perfectly absorb an incident electromagnetic wave if all the wave components scattered out by the device destructively interfere \cite{Salisbury,Fante1988,Gorodetsky1999,Chong2010}.  Here, we use the term absorption in a broad sense, meaning any process by which the energy of an electromagnetic wave is transferred to a medium and hence converted, for example, to heat, electrical current or fluorescence.  Perfect absorption is of great interest for a range of optical applications \cite{Radi2015,Kats2016}, including radar cloaking~\cite{Salisbury,Fante1988,Vinoy1996}, sensing and molecular detection~\cite{Liu2010a, Kravets2013},  photovoltaics~\cite{Luque2008}, and photodetection~\cite{Konstantatos2010, Knight2010}.

A classic example of interference-assisted absorption is the Salisbury screen \cite{Salisbury,Fante1988}, an absorber for microwaves consisting of a thin resistive sheet placed above a flat reflector.  With a specific sheet resistivity and a spacing of exactly one quarter of the wavelength, the reflectance goes to zero due to destructive interference between the scattering pathways in the device; an incident wave therefore delivers all of its energy into the sheet.  Perfect absorption can also be achieved with lossy resonant cavities in the regime in which the cavities act as critically-coupled resonators ~\cite{Gorodetsky1999,Cai2000,Tischler2006} at infrared and optical frequencies.  Another conceptually distinct route to total absorption is ideal impedance matching between free space and a lossy substrate, for example using metamaterials~\cite{Nefedov2013, Baranov2015}.

The Salisbury screen and similar devices are one-port networks: the light enters through a single channel (such as a plane wave incident on one face of a slab with a perfect reflector on the other side) and is perfectly absorbed owing to destructive self-interference. A coherent perfect absorber (CPA) \cite{Chong2010} generalizes this phenomenon to arbitrary incoming waveforms that can consist of two or more waves, such as waves incident on opposite faces of an open slab or film \cite{Chong2010,Wan2011,Zhang2012}.  To achieve perfect absorption, it is essential to appropriately choose the values of the system parameters, the operating wavelength and the input waveform, including the intensities and relative phases of the input beams. The sensitivity of CPAs to the input waveform provides the opportunity to flexibly control light scattering and absorption~\cite{Zhang2012, Fang2015,Zheludev16}.  The possibility of achieving perfect absorption at subwavelength lengthscales is extremely interesting for nanophotonic applications.

In this Review, we discuss the theoretical ideas underlying coherent perfect absorption and overview recent advances in realizing various types of CPAs with several platforms, including ultrathin metasurfaces, graphene-based structures, gain/loss structures and waveguides.  We also discuss emerging applications for CPAs in sensing and all-optical information processing. Although our emphasis is on classical optics, coherent perfect absorption is a general phenomenon arising from the interplay of interference and dissipation, and can therefore be realized in other interesting contexts, including acoustics~\cite{Wei2014,Song2014,Ma2014,Duan2015}, polaritonics~\cite{Zanotto2014}, and quantum optics~\cite{Roger2015}.

\section*{Theoretical background}
\label{sec:theory}

\begin{figure*}
\includegraphics[width=0.7\textwidth]{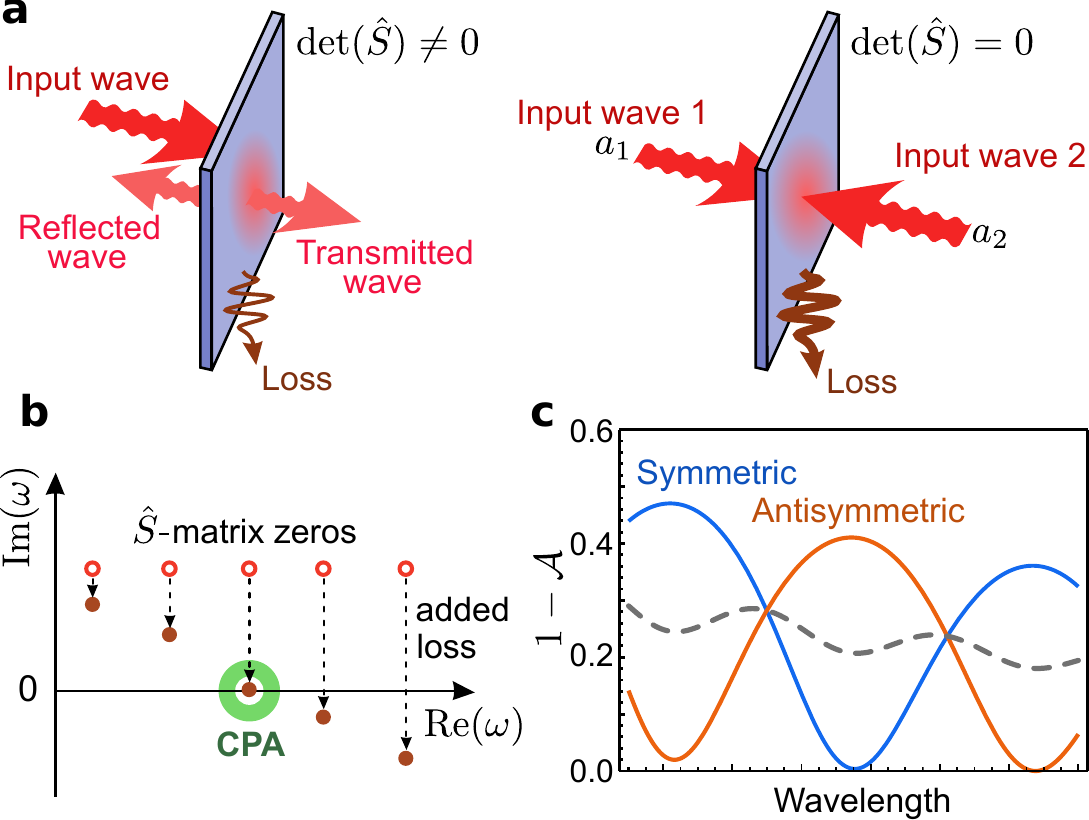}
\caption{\textbf {Conceptual basis of coherent perfect absorption.} (a) Sketch of a 'two-port scatterer', such as a slab or thin film that can be illuminated from either side.  Light incident from one side undergoes partial transmission, reflection and absorption into the material (left).  With coherent beams incident from both sides, the reflected and transmitted waves can destructively interfere on each side, resulting in perfect absorption (right).  (b) Zeros of the $\hat{S}$ matrix in the complex frequency plane for a slab made of a lossless (hollow circles) and lossy (filled circles)  material. (c)  Wavelength dependence of the normalized output intensity $1 - \mathcal{A}$, where $\mathcal{A}$ is the joint absorption defined in Eq.~(\ref{joint_absorption}).  Results are shown for input beams with equal intensity.}
\label{fig1}
\end{figure*}

The simplest yet most illustrative example of a CPA is a two-port linear system.  This can represent a variety of photonic structures, such as an absorbing planar slab or film illuminated on both sides by coherent light beams with fixed incidence angles (Fig.~1a).  Each input beam is partially reflected and partially transmitted; some of its energy can also be absorbed into the medium.  The total outgoing wave on each side, therefore, is a superposition of a reflected wave and of a wave transmitted from the opposite side.  

This picture can be described rigorously using the scattering matrix formalism~\cite{SalehTeich,Haus}.  The relationship between input and output waves is described by
\begin{equation}
  \begin{pmatrix}
    b_1 \\ b_2
  \end{pmatrix} = \hat{S}
  \begin{pmatrix}
    a_1 \\ a_2
  \end{pmatrix}, \;\;\;
  \hat{S} = \begin{pmatrix}
    r_{11} & t_{12} \\ t_{21} & r_{22}
  \end{pmatrix},
\label{eq1}
\end{equation}
where the complex scalars $a_i$ and $b_i$ denote, respectively, input and output wave amplitudes in the $i$-th channel (or port); $r_{ii}$ are the reflection coefficients, and $t_{ij}$ the transmission coefficients.  This formalism can be extended to describe more scattering channels and to include polarization degrees of freedom.  The scattering matrix $\hat S$ is derived from Maxwell's equations and depends on the geometry of the scatterer, material properties and operating wavelength.  It can be calculated in several ways, including the transfer matrix method \cite{SalehTeich}, Green's function methods \cite{Lee1981,Sentenac2013}, finite-element methods with scattering boundary conditions \cite{JinFEM} and coupled-mode theory \cite{Haus,Poladian,Suh2004,Kang2013,Kang2014,Zhu2016}.

Coherent perfect absorption occurs when every output wave component vanishes, i.e.,
\begin{equation}
\hat{S} \, \mathbf{a}_{\textrm{CPA}} = {\mathbf{0}},
\label{eq2}
\end{equation}
where $\mathbf{a}_{\textrm{CPA}}$ is a non-zero vector of input wave amplitudes.  This means that at least one eigenvalue of $\hat{S}$ is zero; the eigenvector $\mathbf{a}_{\textrm{CPA}}$, the 'CPA eigenmode', coincides with a solution to Maxwell's equations obeying purely incoming boundary conditions.  If the structure is supplied with a different set of inputs, not matching a CPA eigenmode, Eq.~(\ref{eq2}) will not be satisfied and outgoing waves will emerge.

  
Symmetries of the scatterer result in symmetries of the $\hat S$ matrix. If the system is Hermitian --- that is, the dielectric function and permeability are  real everywhere --- it follows from Maxwell's equations that $\hat S$ is unitary, making Eq.~(\ref{eq2}) impossible to satisfy. In other words, if the material is not lossy or amplifying, the electromagnetic energy entering the system is equal to that leaving it. For non-magnetic lossy media, the dielectric function $\varepsilon(r)$ has a positive imaginary part, which results in eigenvalues of $\hat S$ smaller than unity. Note that the scattering matrix formalism distinguishes between intrinsic material losses, which are induced by dissipation of electromagnetic energy, and out-coupling losses, which are caused by the openness of the system. For instance, out-coupling losses are present even in the Hermitian case.

In most cases, $\hat{S}$ is also constrained by optical reciprocity~\cite{Reciprocity}.  In the steady-state and linear regime and in the absence of magneto--optical materials, $\hat{S}$ must be symmetric; for the two-port case, this means that $t_{12}=t_{21}=t$. Importantly, adding loss (or gain) does not necessarly violate reciprocity, despite breaking time-reversal symmetry. Unless otherwise specified, in the following we assume that reciprocity holds.

The $\hat{S}$ matrix can also be constrained by geometrical symmetries.  For instance, if a two-port structure is symmetric under a mirror reflection that exchanges the ports, $ r_{11}=r_{22}=r$. The two eigenvalues of $\hat{S}$ are then $r \pm t$ and the corresponding eigenvectors $\mathbf{a}_\pm \propto \left( {1, \pm 1} \right)$, with the $+$ and $-$ solution corresponding to a symmetric and antisymmetric input of equal intensity.


To understand how the CPA condition set by Eq.~(\ref{eq2}) can be satisfied, we analytically continue the frequency $\omega$ (on which $\hat{S}$ depends) to complex values \cite{Chong2010}.  Zeros of $\hat{S}$ --- that is, occurrences of zero eigenvalues --- generally occur at discrete points in the complex $\omega$ plane (Fig.~1b).  When $\varepsilon(r)$ is real, the zeros lie in the upper half-plane, $\textrm{Im}[\omega]>0$; these have a one-to-one correspondence with the scattering resonances of the system, and the zeros themselves coincide with 'unphysical' solutions featuring input waves that grow exponentially with time~\cite{Chong2010,Weinstein}.  If loss is increased by adding a positive imaginary part to $\varepsilon(r)$, the zeros move in the plane, typically downwards.  For a certain amount of loss, a zero can move onto the real axis, $\rm Im [\omega]=0$, producing a physical CPA eigenmode.

The CPA condition can be interpreted as the time-reverse of the lasing threshold \cite{Poladian,Chong2010}.  The complex conjugate of a CPA solution, for real $\omega$, is a wave that is purely outgoing rather than incoming.  Such a solution describes a laser poised exactly at the lasing threshold~\cite{Haken,SALT}; it satisfies Maxwell's equations for a dielectric function $\varepsilon^*(r)$ that exhibits gain rather than loss (note that above the lasing threshold, the dielectric function becomes intrinsically nonlinear and the CPA-laser correspondence typically breaks down \cite{Reddy2013a,Reddy2013b}). In the complex $\omega$ scenario, threshold lasing solutions correspond to poles, rather than zeros, of $\hat{S}$ (that is, points at which an eigenvalue of $\hat{S}$ diverges).  For real $\varepsilon(r)$, the poles lie in the lower half-plane, mirroring the CPA zeros; adding gain moves the poles and zeros upward.  This relationship between zeros and poles is particularly important for the CPA-laser structures discussed later. Note that this is unrelated to the idea of laser cooling as an 'optically pumped laser running in reverse'~\cite{lasercooling}.

Absorption in a CPA is highly sensitive to the illumination conditions, particularly to the relative phase of the input field components.  For a two-port structure, we can define the joint absorption~\cite{Baldacci2015} as
\begin{equation}
  \mathcal{A} \equiv
  1 - \frac{\left|b_1\right|^2 + \left|b_2\right|^2}
           {\left|a_1\right|^2 + \left|a_2\right|^2}.
           \label{joint_absorption}
\end{equation}
This is a convenient figure of merit, because $\mathcal{A} = 0$ for a Hermitian structure and $\mathcal{A}=1$ for a CPA.  The normalized output intensity  $1-\mathcal{A}$ for a uniform dielectric slab near a CPA frequency is plotted in Fig.~1c for symmetric, antisymmetric and incoherent input beams of equal intensity. CPAs may occur both for symmetric or antisymmetric modes, depending on the frequency. Remarkably, at the CPA frequency of the symmetric and antisymmetric mode, the other eigenmode of $\hat{S}$ experiences a lower absorption than incoherent inputs: this is because the outgoing wave components interfere constructively, suppressing absorption.  This analysis can be extended to two-port structures lacking mirror symmetry, whose CPA eigenmodes consist of beams with unequal intensities \cite{Piper2014a, Piper2014, Baldacci2015, Graphene2017}.

An important limiting case of a two-port CPA is a flat resistive sheet much thinner than the operating wavelength, such as a metasurface or a graphene layer.  For normal incidence, $\hat{S}$ has a simple form. In addition to reciprocity and mirror symmetry, the reflection coefficient $r$ and the transmission coefficient $t$ are expressed through the complex-valued sheet conductivity $\sigma_s$ (in SI units), as
\begin{equation}
  r = -\left(1+\frac{2\varepsilon_0 c}{\sigma_s}\right)^{-1}, \;\; t = r + 1,
  \label{thinsmatrix}
\end{equation}
where $\varepsilon_0$ is the permittivity of free space and $c$ is the speed of light~\cite{Radi2015,Luo2014,Liu2014}. In the absence of gain, energy conservation requires that
\begin{equation}
|r|^2+|r+1|^2\leq 1.
\label{sheet}
\end{equation}
From this, it can be shown that absorption under single-port illumination can only reach a maximum of 50\%, which occurs when $\sigma_s \,=\, 2\varepsilon_0 c \,\approx\, 1/(188.4\,\Omega)$ and $r=-1/2$. It follows from Eq.~(\ref{thinsmatrix}) that the eigenvalues of $\hat{S}$ are $\{-1, 1+2r\}$, thus satisfying $r=-1/2$ realizes a CPA: if the sheet is illuminated symmetrically with two coherent beams, it coincides with an anti-node of the electric field and absorption reaches 100\%~\cite{Zhang2012,Radi2015,Li2015}.  Conversely, for antisymmetric beams, the sheet is located at a node and the absorption vanishes.  This example demonstrates that perfect absorption can occur even for an absorber of subwavelength thickness; this can be generalized to more complex 2D and 3D geometries, and is one of the most interesting features of CPAs.

Perfect absorption at subwavelength thickness also underlies the operation of the already mentioned Salisbury screen \cite{Salisbury,Fante1988}.  The separation between the  resistive sheet and the perfect reflector  ensures that the incident wave and the wave returning from the reflector enter the sheet with equal phase; if the sheet conductivity is tuned to $\sigma_s \,=\, 2\varepsilon_0 c$, perfect absorption ensues.  Interestingly, changing the reflector type (for instance, from a perfect electrical conductor to a perfect magnetic conductor) affects how the incident and reflected waves interfere at the absorbing region, changing the total absorption.  Single-port thin-film absorbers and their relationships with CPAs are discussed in detail in Refs.~\cite{Li2014,Li2015,Radi2015,Kats2016}.  It is worth emphasizing that one-port devices cannot exhibit coherent control of absorption through beam modulation.


\section*{Planar structures}
\label{sec:planar}
\subsection*{Dielectric slabs}
\label{sec:slabs}

\begin{figure*}[!t]
\includegraphics[width=0.7\textwidth]{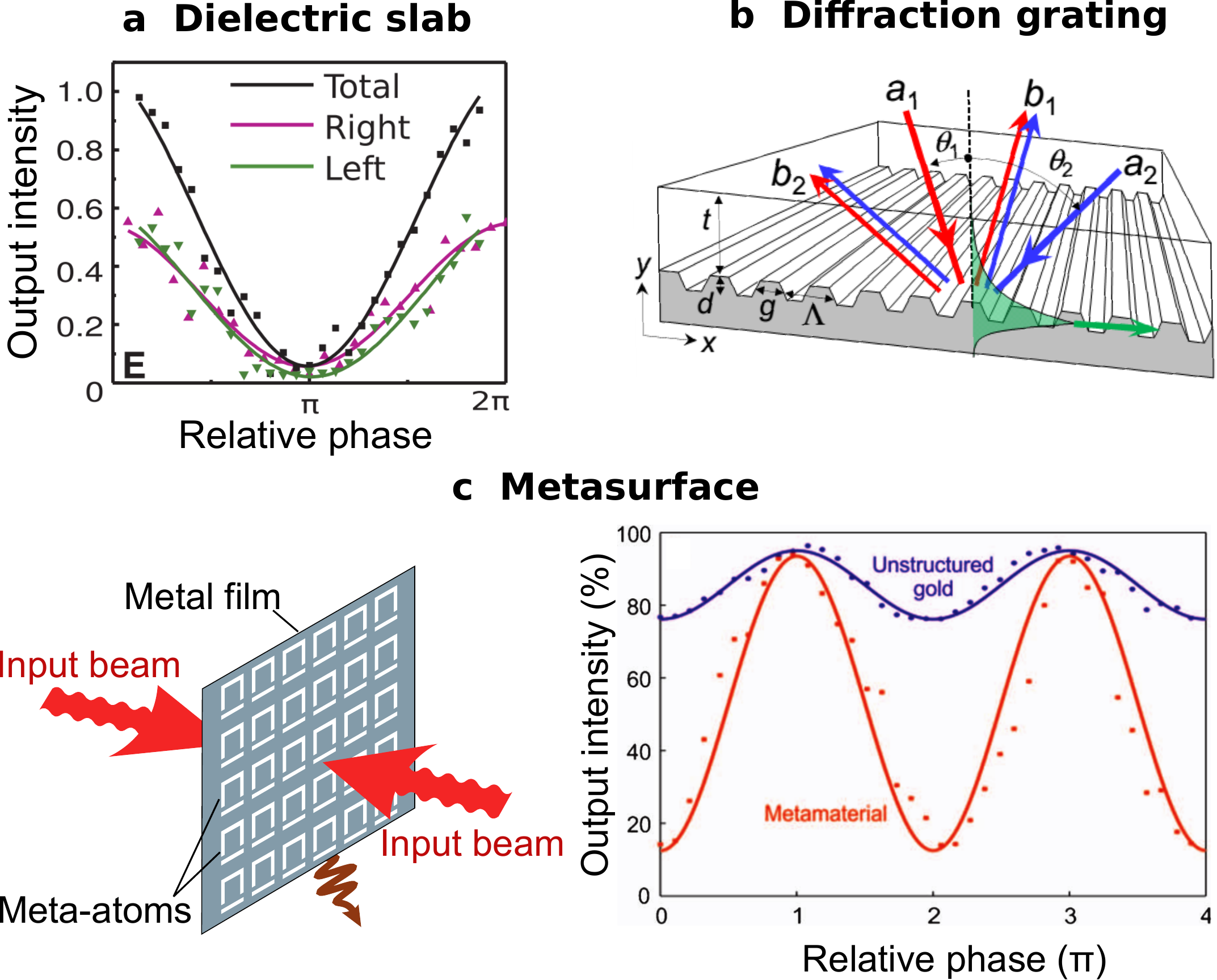}
\caption{\textbf {Realization of CPAs in planar structures.} (a) Experimental results showing coherent perfect absorption  in a silicon wafer of thickness $\sim 110\,\mu\textrm{m}$ illuminated by normally incident beams  with wavelength $\lambda = 998.4\,\textrm{nm}$ and equal intensity.  The total output intensity, and the intensities emitted to the left and right of the sample, are plotted as a function of the relative phase of the input beams.   (b) Schematic depiction of a coherent perfect absorber (CPA) based on a plasmonic (silver) diffraction grating; two coherent beams with amplitudes $a_{1}$ and $a_{2}$ (corresponding to different diffraction orders) are incident on a lossy grating with angles $\theta_1$ and $\theta_2$, and the outputs $b_1$ and $b_2$ can vanish.  (c)  Schematic representation of an ultrathin resonant metasurface, which acts as a CPA when placed at an anti-node of a standing wave, accompanied by the measured output intensities.  Panel a is reprinted with permission from Ref.~\citenum{Wan2011},  panel b is reprinted with permission from Ref.~\citenum{Song_2015}, the right part of panel b  is reprinted with permission from Ref.~\citenum{Zhang2012}.}
\label{planar}
\end{figure*}

The simplest realization of a CPA is a slab of lossy homogeneous material illuminated by plane waves from both sides~\cite{Chong2010,Longhi2010a,Gmachl2010}, which acts as the time-reverse of a Fabry-P\'erot laser \cite{Haus}.  A very similar effect was also discussed in the context of optical fiber gratings~\cite{Poladian} (the CPA eigenmodes were called 'loss resonances' and identified as the time-reverse of threshold lasing modes).

A simple experimental demonstration of a slab CPA was realized  using a silicon wafer~\cite{Wan2011}.  CPA resonances were observed in the $1000$~nm wavelength range, with measured output intensities showing strong dependence on the relative phase of the two input beams.  The intensities approached zero at $\pi$ phase difference (Fig.~2a).  Several other planar CPAs have been theoretically studied, including composite metal--dielectric films~\cite{Dutta-Gupta2012a}, heavily-doped silicon films~\cite{Pu2012} and homogeneous lossy slabs bounded by dispersive mirrors~\cite{ABOURADDY15}.

It was predicted that dielectric photonic crystal slabs can realize a variant of perfect absorption known as 'degenerate critical coupling' \cite{Piper2014a, Piper2014}.  Typically, the zeros of $\hat{S}$ occur at distinct frequencies; however, it is possible to design a resonant two-port mirror-symmetric cavity supporting degenerate resonances with opposite mirror symmetries, so that both eigenvalues of $\hat{S}$ are zero at a certain frequency.  The absorption at that frequency becomes insensitive to the relative phase of the two input beams --- the exact opposite of the absorption modulation discussed above.  Moreover, perfect absorption occurs regardless of the intensity of the input beams, and even for a single input beam incident on either port.  This phenomenon has been demonstrated numerically in a photonic crystal slab incorporating a graphene layer to provide dissipation; the slab contained guided resonances that could be tuned by changing the crystal parameters to achieve the mode degeneracy \cite{Piper2014a, Piper2014}.

\subsection*{Diffraction gratings}
\label{sec:gratings}

Another route to realizing multi-port CPAs is to use a diffraction grating \cite{Yoon2012,Dutta-Gupta2012,Song_2015}.  A grating of period $D$ couples diffraction orders with lateral wavenumbers differing by $2\pi/D$.  The input channels are two beams incident on the same side of the grating, but at different angles (Fig.~2b).  The angles are chosen so that the lateral wavenumbers match the grating's surface plasmon polariton mode at different diffraction orders.  This produces a $2\times 2$ scattering problem in which  the input and output channels are not reciprocal partners; the effective $\hat{S}$ matrix is asymmetric  (the actual reciprocal partners do not satisfy the wavenumber-matching condition, and are thus decoupled).  Because the absorption is mediated by surface plasmons, this CPA can be regarded as the time-reverse of a spaser \cite{Bergman}, and is accompanied by a substantial field enhancement around the grating~\cite{Yoon2012}.

The realization of a CPA based on the diffraction grating configuration has been reported in REF.~\cite{Song_2015}.  The tuning of the system to the CPA condition can be aided by incorporating a medium with adjustable optical gain~\cite{Giese2014,Yoon2015,Song_2015} or by varying the temperature (which alters the free-electron collision rate in a metal grating~\cite{Song_2015}).

\subsection*{Metasurfaces}
\label{sec:metasurfaces}

There is currently a considerable interest in using planar metamaterials --- metasurfaces --- to realize CPAs~\cite{Zhang2012, Kang2013, Fang2014, Kang2014, Nie2014, Fang2015, Fang2016, Zhu2016}.  The meta-atoms (unit cells of the metasurface) can be designed to tailor the electromagnetic properties at any desired frequency~\cite{Meinzer}.  A metasurface-based CPA was first demonstrated  using a free-standing plasmonic metasurface (Fig.~2c) ~\cite{Zhang2012}, which had thickness $\lambda/13$ ($\lambda$ indicates the wavelength) and was tailored to function as a CPA at exactly $\lambda = 1550\,\textrm{nm}$ with normally-incident beams on each side.  This system acts like the thin-film CPA described in Eq.~(\ref{sheet}), with the meta-atom parameters determining the effective sheet conductivity.  Subsequent experiments demonstrated ultrafast switching between the absorbing and transparent regimes of a metasurface, obtained by controlling the relative input phase~\cite{Fang2014}. 

A polarization-independent metasurface CPA was designed using dipolar meta-atoms \cite{Kang2013}. The meta-atoms were optimized with the aid of coupled-mode theory, a flexible technique that can be applied to other CPAs~\cite{Kang2014}. Other theoretical proposals for metasurface CPAs include a bilayer asymmetric split-ring meta-atom design~\cite{Nie2014}, ceramic fishnet structures for operation in the microwave regime~\cite{Zhu2016} and a self-complementary checkerboard-like metasurface for terahertz waves~\cite{Kitano2016}.  Owing to the design flexibility offered by metasurfaces, coherent excitation by multiple beams can be used to achieve a variety of other interesting effects apart from perfect absorption.  These include the control of chirality and birefringence \cite{Zh-2015, Zh-2014-APL}, anomalous refraction \cite{Zh-2014-OE} and excitation of high-order resonances \cite{Zh-2016, Zh-2017b}.

The absorption bandwidth of metasurface CPAs depends on their design.  Metasurfaces with strongly resonating meta-atoms maintain their desired properties over a narrow bandwidth, whereas weakly resonating metasurface CPAs  can exhibit broad absorption bandwidths, similar to the graphene-based CPAs discussed in the next section.

\subsection*{Graphene-based structures}
\label{sec:graphene}

Another viable method to realize CPAs, particularly at terahertz frequencies, is to use graphene as the absorbing medium~\cite{Liu2014, Fan2014, Rao2014, Fan2015, Hu_JN_2016, Graphene2017}. Coherent control of absorption has been experimentally demonstrated for a 30-layer graphene film sandwiched between silica substrates, with strong modulation (80\%) observed in the absorption~\cite{Rao2014}.  CPAs may also provide a convenient way to access optical nonlinearities in graphene, such as four-wave mixing processes~\cite{Rao2014,Rao2015}.

A single sheet of undoped graphene exhibits a frequency-independent single-beam absorption of 2.3\%, which is insufficient for a thin-film CPA.  There are various methods to reach the required 50\% single-beam absorption, such as increasing the carrier concentration through chemical doping~\cite{Koppens2011,Fan2014}, applying an electrostatic potential to change the Fermi level~\cite{Mak2008, Wang2008, Koppens2011, Polini2012, Liu2014}, patterning the graphene sheet to induce local resonances~\cite{Fan2015,GrapheneMeta} or using multilayer graphene \cite{Rao2014,Graphene2017}.

Single-layer graphene doped to around 200~meV (either chemically or electrostatically), exhibits 50\% single-beam absorption at terahertz frequencies~\cite{Liu2014}.  Electrostatic doping is appealing because it is easily tunable~\cite{Liu2014,Fan2015}, but requires a substrate.  For free-standing single-layer graphene CPAs \cite{Fan2014}, chemical doping is required.  In either case, the sheet conductivity for unpatterned graphene is described by the Drude formula $\sigma(\omega) = \sigma(0)/(1-i\omega\tau)$, where $\tau$ is the DC transport scattering time.  Because $\sigma(\omega)$ lacks resonant features, the resulting CPA frequency bandwidth can be very large. Graphene can also be combined with a metasurface in a single CPA device~\cite{Hu_JN_2016}. 

\subsection*{Broadband and narrowband operation}
\label{sec:bandwidth}

The bandwidth of a CPA is the frequency range over which absorption remains nearly perfect.  A broad bandwidth is desirable for many telecommunications-related applications, such as optical modulators and logic elements.  As discussed, zeros of $\hat{S}$ correspond to resonances; for a uniform dielectric slab, these are Fabry-P\'erot resonances, thus the CPA bandwidth is the Fabry-P\'erot free spectral range $\Delta\omega = \pi/nL$, where $n$ is the refractive index and $L$ is the slab thickness \cite{Chong2010,Chong2013}.

In the thin-film CPA limit, $L \rightarrow 0$, the CPA bandwidth can be very high, and is limited principally by the frequency dispersion of the sheet conductivity $\sigma_s$.  This has been demonstrated experimentally at microwave frequencies (6--8\,GHz) using a thin ($L \sim 10^{-4}\lambda$) conductive film~\cite{Li2015}.  High bandwidths are a compelling feature of graphene-based CPAs. 

Apart from using thin films, there are several other methods to boost the CPA bandwidth.  One is based on white-light cavities~\cite{Xiao_2008}, which use superluminal phase elements to compensate the phase accumulated by light propagating through the cavity; it has been predicted that an absorbing layer surrounded by two white-light cavities can act as a CPA for an operating wavelength of  1.5~$\mu$m, with a bandwitdth of $\approx 40\,\textrm{nm}$ ~\cite{Kotlicki2014}.  Another design for a broad-band CPA uses epsilon-near-zero (ENZ) materials, which are materials with permittivity close to zero~\cite{Kim2016a}.  At the ENZ material interface, the normal component of the electric field obeys the boundary condition ${\varepsilon _1}E_1^\bot={\varepsilon _{{\rm{ENZ}}}}E_{{\rm{ENZ}}}^\bot$. Therefore, the electric field can become very large in the ENZ material, enabling strong light absorption.  Finally, broadband CPA in a full octave (800--1600 nm) has been experimentally realized by tailoring aperiodic dielectric mirrors on both sides of a silicon cavity~\cite{ABOURADDY16}.

Very narrow-band CPAs might also be useful for certain applications, such as sensors and narrow-band optical switches~\cite{Darkness}.  One strategy for achieving a narrow-band CPA exploits the ability of a Fano-resonant system~\cite{Miroshnichenko2010, Lukyanchuk2010} to host narrow asymmetric resonances~\cite{Yu2015}. This idea has been explored through simulations of a photonic crystal cavity waveguide system; the bandwidth of the resulting CPA was found to be on the order of one hundredth of the operating frequency~\cite{Yu2015}.

Another important property is the range of incident angles over which a system exhibits near-perfect absorption.  This is particularly important when the incident illumination is non-planar, as for light emitted by nearby point sources.  It was shown that light from two coherent point sources can be completely absorbed by locating a small sink in a slab of material with negative refractive index~\cite{Klimov2012}.  A more advanced approach is to design a nonlocal transversely homogeneous multilayer planar slab~\cite{PTlens}; such a structure operates as an omnidirectional CPA, able to perfectly absorb light emitted by coherent point sources  located symmetrically on both sides of the slab.

\subsection*{Polarization effects}

Although many CPA designs operate at a fixed linear polarization or respond equally to any incident polarization~\cite{Kang2013}, the concept can be extended to structures with nontrivial polarization characteristics.  For instance, a chiral slab can have separate $\hat{S}$ matrices for clockwise and counterclockwise polarization states, with the CPA conditions for  clockwise and  counterclockwise modes satisfied at different frequencies~\cite{Ye2016}.  The CPA eigenmodes then consist of input waves of the same circular polarization but different intensities, in contrast to the non-chiral case, in which CPA eigenmodes consist of two beams of equal intensities.  Chiral CPA structures may have applications in coherent polarization control~\cite{Kang2015,Ye2016}.

An unusual phenomenon known as 'polarization standing waves' arises in CPA-like systems with incident circularly-polarized waves~\cite{Fang2016}.  Unlike conventional energy standing waves, polarization standing waves have constant electromagnetic energy densities, but a periodically modulated polarization state along the axial direction. Light absorption in an anisotropic sample becomes dependent on the sample orientation relative to the polarization. On the other hand, the absorption is insensitive to the sample's axial position, in sharp contrast to thin-film absorption in a conventional energy standing wave, which depends on the axial position of the film relative to the nodes and antinodes of the wave.

\section*{$\mathcal{PT}$-symmetric structures}
\label{sec:PTsymmetry}

\begin{figure*}
\includegraphics[width=.8\textwidth]{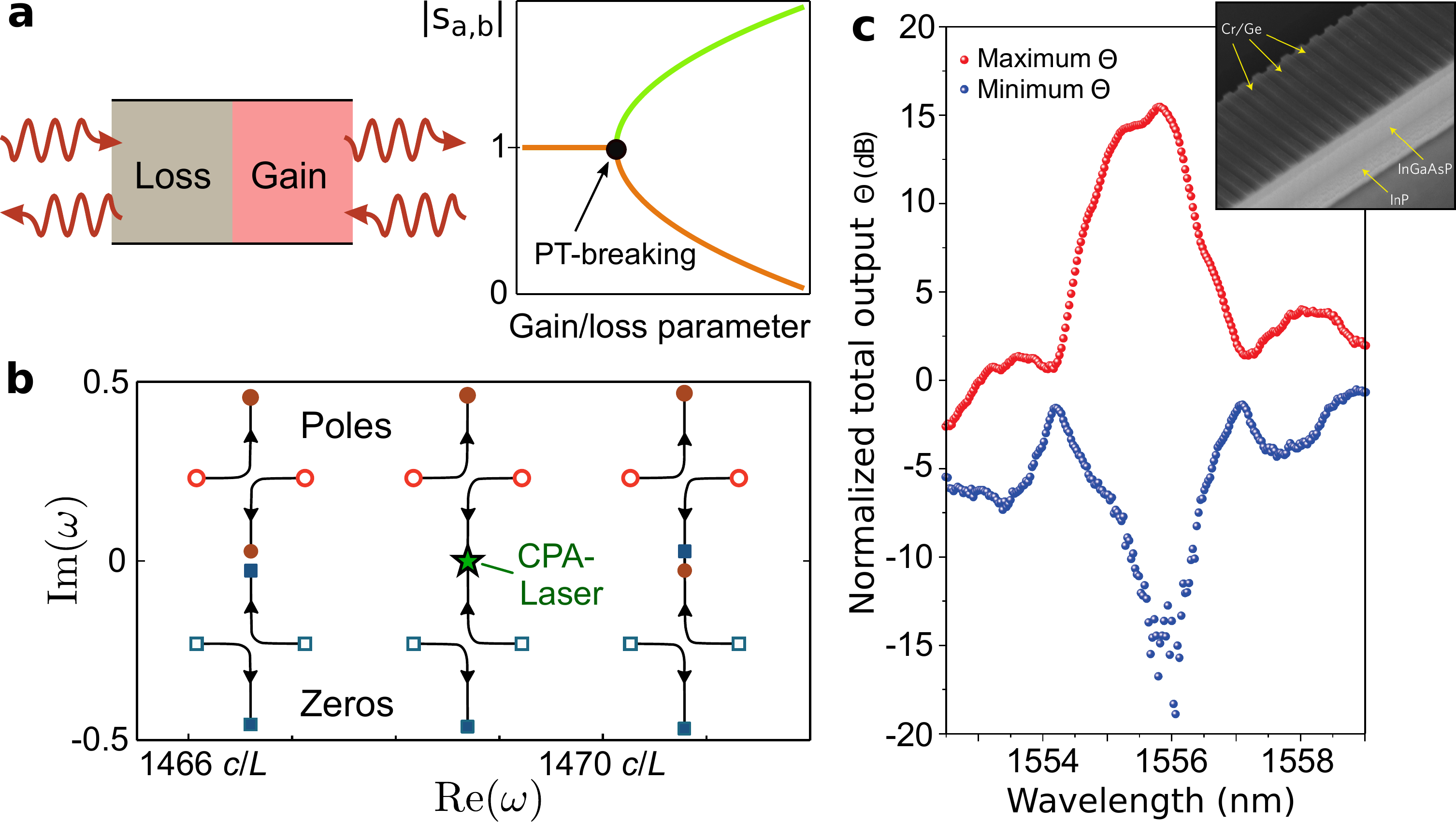}
\caption{\textbf {Coherent perfect absorption in parity- and time-reversal ($\mathcal{PT}$) symmetric structures.} (a) Schematic depiction of a two-port $\mathcal{PT}$-symmetric gain/loss bilayer; $\mathcal{PT}$ breaking is manifested as a bifurcation of the $\hat{S}$ matrix eigenvalues ($s_a$ and $s_b$) as the  level of balanced gain/loss changes, as shown on the graph. (b) Trajectories in the complex frequency plane for the $\hat{S}$ matrix zeros (circles) and poles (squares) of a $\mathcal{PT}$-symmetric bilayer (as the one in panel a) as the gain/loss level is changed. All curves start with zero gain and loss; gain and loss increase proportionally along the trajectories.  The  coherent perfect absorber (CPA)-laser behaviour occurs when a zero and a pole meet on the real axis. (c) Experimental results showing CPA-laser functionality.  The device consists of a waveguide on an InP substrate, with InGaAsP quantum wells and Cr/Ge absorbers serving as periodic gain and loss layers.  At the 1556\,nm operating wavelength, the normalized output varies between 15 dB (laser) and -15 dB (CPA), depending on the relative phase of the two inputs.  {\it c}, speed of light. Panel b is  reprinted with permission from Ref.~\citenum{Chong2011}, panel c is reprinted with permission from Ref.~\citenum{Wong2016}.}
\label{PT}
\end{figure*}

Parity- and time-reversal ($\mathcal{PT}$) symmetric optics has attracted considerable interest~\cite{Bender1998, Bender2002, Makris2008, Guo2009, Ruter2010}.  Among other intriguing properties, $\mathcal{PT}$-symmetric structures can be used to realize CPAs with unique features. The concept of $\mathcal{PT}$ symmetry originated in the theoretical quantum physics literature~\cite{Bender1998, Bender2002}, in which a Hamiltonian is said to be $\mathcal{PT}$-symmetric if it is invariant under consecutive time ($\mathcal{T}$) and parity ($\mathcal{P}$) reversal operations~\cite{Bender1998, Bender2002, UFN}.  The concept can be extended to electromagnetics \cite{Makris2008, Guo2009, Ruter2010}, where time reversal interchanges gain and loss.  An optical structure is $\mathcal{PT}$-symmetric if the dielectric function satisfies the condition
\begin{equation}
\varepsilon ({\bf{r}}) = {\varepsilon ^*}(\mathcal{P}\,{\bf{r}}),
\end{equation}
where $\mathcal{P}$ is a parity operation (for example a mirror reflection). An example of such a system are adjacent gain/loss layers (Fig.~3a). Despite being non-Hermitian, $\mathcal{PT}$-symmetric systems exhibit features similar those of Hermitian systems; for instance, they can have real eigenfrequencies within a parameter domain called the $\mathcal{PT}$-unbroken phase.  They can also undergo symmetry-breaking transitions, causing the eigenvalues to coalesce and become complex~\cite{Bender1998, Bender2002, Makris2008, Guo2009, Ruter2010}.

A one-dimensional $\mathcal{PT}$-symmetric structure can act as a CPA and threshold laser simultaneously~\cite{Longhi2010, Chong2011,Longhi14}.  If the system is not separately $\mathcal{P}$- and $\mathcal{T}$- symmetric, $\hat{S}$ is not unitary.   However, $\hat{S}$ can exhibit a $\mathcal{PT}$-unbroken phase in which its eigenvalues $\{s_i\}$ are unimodular, $|s_i|=1$, implying a net conservation of energy.  Alternatively, the eigenvalues can pair up and undergo $\mathcal{PT}$-symmetry breaking, such that the eigenvalues $s_a$ and $s_b$ are not unimodular, but satisfy $s_a s_b^* = 1$ (Fig.~3a). Effectively, one eigenmode is amplified and its partner correspondingly damped by the system.  If an eigenvalue goes to zero ($s_a=0$), indicating a CPA, the other eigenvalue simultaneously becomes infinite ($s_b=1/s_a=\infty$), indicating threshold lasing.  This is illustrated  for a two-port gain/loss bilayer in Fig.~3b.  Notably, CPA and lasing are allowed even though the underlying structure has zero net gain or loss.

Coherent perfect absorption in a system with an effective $\mathcal{PT}$ symmetry was first demonstrated experimentally using a pair of coupled microwave resonators~\cite{Sun2014}.  Recently, two independent papers reported the realization of $\mathcal{PT}$-symmetric laser-absorbers based on optical-frequency waveguides~\cite{Gu2016,Wong2016}.  In the experiments described in Ref.~\citenum{Gu2016}, a halving of the emission peaks was observed, indicating the occurrence of the $\mathcal{PT}$-broken phase, although there was no direct evidence for CPA-laser behaviour.  For the system discussed in Ref.~\citenum{Wong2016}, the input and output intensities were directly measured, and for different phase offsets of the inputs both a maximum and a minimum of the relative output intensity were observed, with a contrast of around 30\,dB (Fig.~3c).  The pump intensity was maintained near the lasing threshold to prevent above-threshold gain nonlinearity from distorting the $\mathcal{PT}$ symmetry.

Other $\mathcal{PT}$-symmetric CPA-laser designs have been theoretically studied, including adjacent atomic cells filled with $\Lambda$-type three-level atoms~\cite{Hang2016}, plasmonic phase-controlled absorber-amplifier systems~\cite{Dione} and non-$\mathcal{PT}$-symmetric systems with purely imaginary effective permittivity and permeability \cite{PTzero}.

\section*{Guided mode structures}
\label{sec:guided}

\begin{figure*}
\includegraphics[width=1\textwidth]{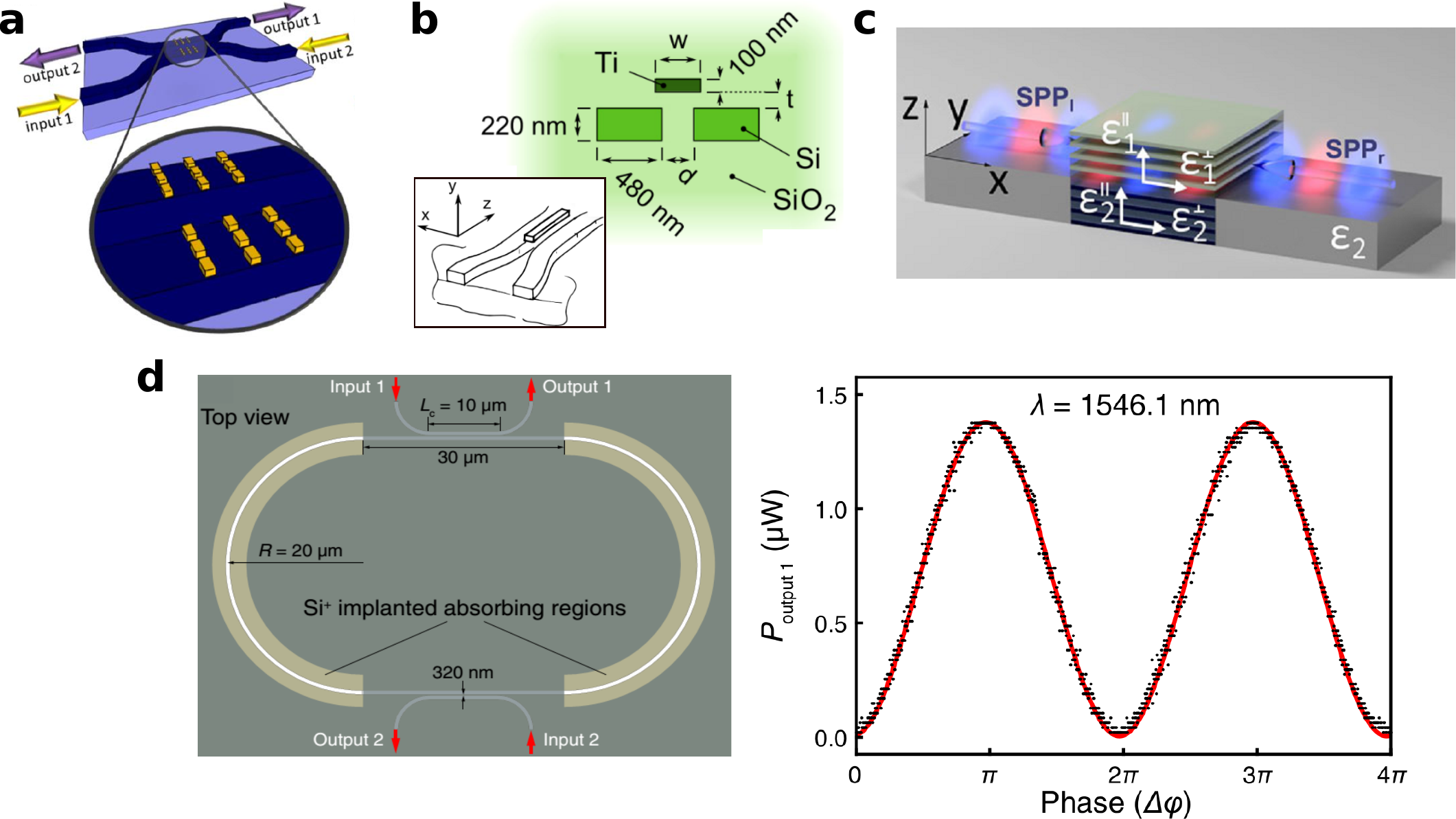}
\caption{\textbf {Coherent perfect absorption in guided mode structures.} (a)~Schematic illustration of two coupled silicon-on-insulator waveguides with plasmonic nanoantennas placed on the surface of the waveguides, in the middle section of an X-coupler that couples the input and output channels.  (b) Schematic representation of a waveguide coherent perfect absorber (CPA) realized through the integration of a lossy metal component with silicon photonic coupled waveguides~\cite{Zanotto2016}. (c)~Schematic illustration of a surface plasmon CPA free of parasitic out-of-plane scattering; two identical metal-dielectric interfaces form the external region, from which two coherent surface plasmons impinge on an anisotropic absorber formed by layered metal-dielectric metamaterials. The incident plasmons form a standing wave inside the structure, and the energy is absorbed without scattering. (d) Schematic depiction of a silicon racetrack resonator serving as a two-channel CPA, accompanied by the measured output power as a function of the relative phase of the inputs at the CPA wavelength.  Panel a is reprinted with permission from Ref.~\citenum{Bruck}, panel c is reprinted with permission from Ref.~\citenum{Ignatov2016}, panel d is reprinted with permission from Ref.~\citenum{Rothenberg2016}. }
\label{guided}
\end{figure*}

For many applications, it is desirable to realize CPAs in guided mode configurations, such as optical fibers or plasmonic waveguides. These CPAs could serve as slow-light waveguide couplers~\cite{Gutman2013} or as components in integrated optical circuits for signal processing and computing.  In one possible implementation, resonant absorbing nanoantennas are placed within or near the waveguide (Fig.~4a). When two counter-propagating modes excite the nanoantennas, the transmitted waves destructively interfere, increasing the total absorption. If the nanoantennas are designed so that the single-beam absorption is 50\%, the structure can serve as a CPA. This approach has been numerically demonstrated for silicon-on-insulator waveguides at the telecom wavelength~\cite{Bruck} and for metal-insulator-metal plasmonic waveguides in the near-infrared regime~\cite{Park2015}.

An alternative technique to coherently control absorption in guided mode systems is to exploit the near-field interaction between a waveguide and a dissipative coupler (Fig.~4b). The coupler can be a ring resonator, a 1D photonic crystal cavity~\cite{Grote2013} or simply a plasmonic element coupled to the waveguide~\cite{Zanotto2016}. Also in this case, the light interference within the lossy component can increase the absorption, producing a CPA if the parameters are appropriately tuned.

In reality, whenever a waveguide is coupled to a nanoantenna (or any other element that breaks the waveguide's translational symmetry), part of the loss can consist of off-axis scattering into free space modes~\cite{Stegeman1981}.  This is undesirable for applications in which absorption into the medium serves a purpose, such as photocurrent generation.  One solution was theoretically  proposed for a metal-insulator plasmonic waveguide (Fig.~4c): it was shown that a scattering-free coherent surface plasmon absorber may be realized at the interface between two uniaxial media with specific dielectric tensors \cite{Ignatov2016}. Parasitic scattering of surface plasmons into free-space modes is eliminated by the extreme anisotropy of the absorbing layer, leading to ideal matching of plasmon fields in the two regions~\cite{Elser2008}. The incident plasmons are reflected and refracted at the boundary, similarly to plane waves, without out-of-plane scattering.

Advances towards the use of CPAs in integrated nanophotonic structures have recently been reported~\cite{Rothenberg2016, Intracavity}.  A CPA was realized using a photonic silicon racetrack resonator coupled to four input/output ports~\cite{Rothenberg2016} (Fig.~4d): absorption was achieved by Si-ion implantation, and an extinction of almost 25~dB was measured at the 1.54~$\mu$m operating wavelength.  Similar results were obtained for a ring cavity with an integrated variable coupler~\cite{Intracavity}.

\section*{2D, 3D and more complex structures}
\label{sec:complex}

It should also be possible to achieve coherent perfect absorption in 2D and 3D geometries, such as spheres or rods suspended in free space.  The main difficulty is that the CPA eigenmodes are not, in general, planar; for instance, the CPA eigenmodes of a sphere are spherical waves.  The theoretical conditions for realizing plasmonic nanoparticle and nanorod CPAs were investigated,~\cite{Noh2012, Noh2013} and it was found that for a metallic nanosphere with radius far below the free-space wavelength, CPA eigenmodes exist only for incident transverse magnetic spherical waves, and coincide with plasmonic resonances.  Under irradiation with a CPA eigenmode, perfect absorption is accompanied by strong field enhancement near the nanoparticle surface, resulting from perfect coupling of the incident light to a lossy surface plasmon.

Coherent perfect absorption of a partial spherical harmonic by a core-shell nanoparticle has also been studied~\cite{Mostafazadeh2012}.  Interestingly, the requirement for a highly coherent spherical wave input may be loosened with the use of multilayer nanoparticles, allowing for efficient absorption of random wavefronts~\cite{Bai2016}.  For non-spherical nanoparticles in a complex environment, the CPA eigenmode can be derived using Green's tensor techniques~\cite{Sentenac2013}.

In complex structures with a large number of coupled scattering channels, very strong coherent absorption enhancement may be possible, even if perfect absorption is not achieved.  An example is a random structure containing a strongly scattering (optically diffusive) medium with weak material losses, or a network of coupled optical cavities~\cite{Kottos16}.  The system may appear white under ordinary illumination, but nearly-perfect absorption can occur if the multi-channel inputs are optimized so that they form an appropriate complex wavefront~\cite{Chong2011a}.  However, the achievable total absorption strongly degrades if not all available scattering channels can be precisely controlled \cite{Goetschy2013}.

\section*{Quantum CPAs}
\label{sec:quantum}

\begin{figure*}[!t]
\includegraphics[width=0.8\textwidth]{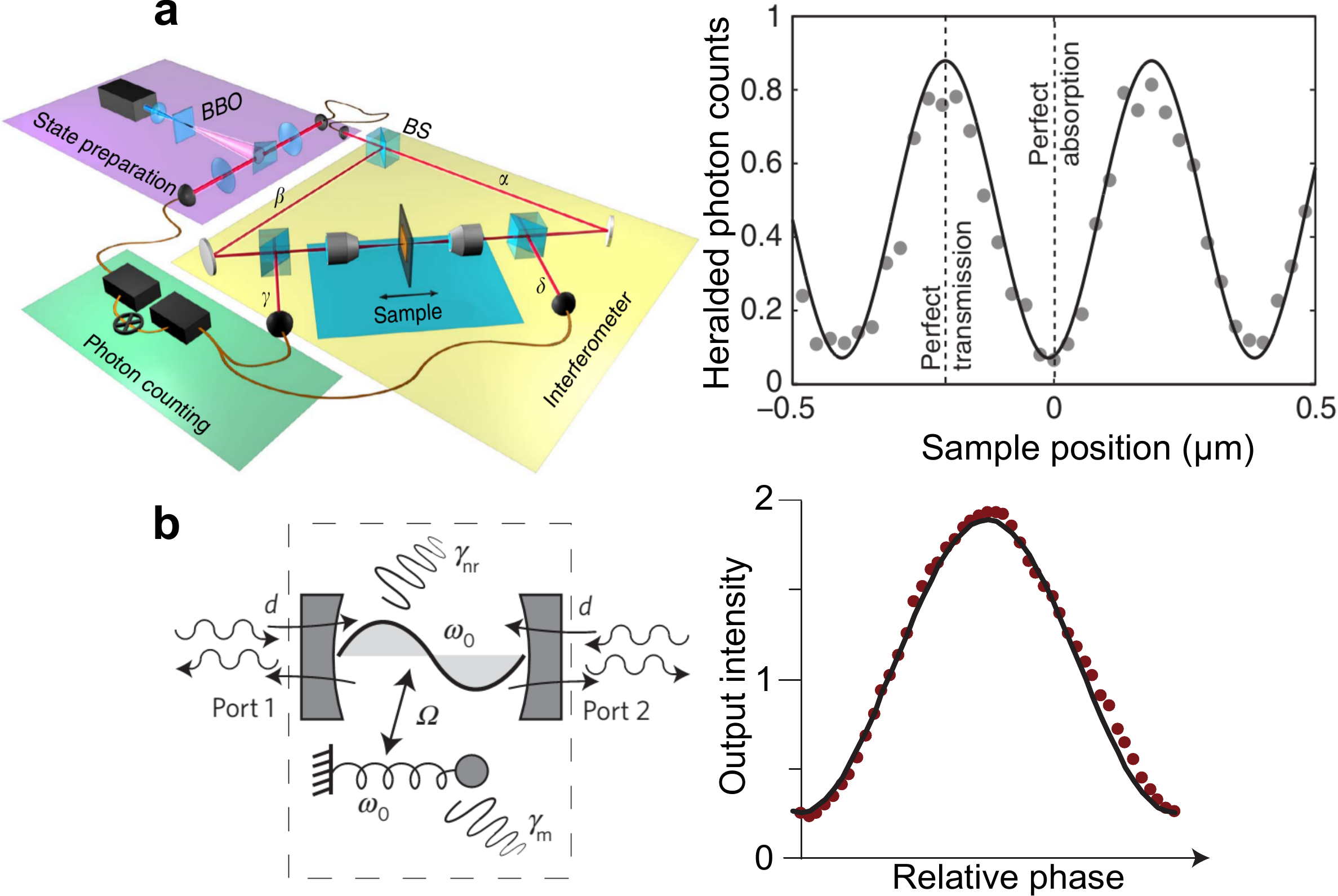}
\caption{\textbf{Coherent perfect absorption in the quantum regime.} (a) A single-photon coherent perfect absorption experiment, using a metasurface coherent perfect absorber (CPA) with inputs prepared using a single-photon source and a beam splitter.  The measured photon counts show that coherent perfect absorption can be obtained also in the single-photon regime.    (b) Coupled-oscillator model for a CPA in the polaritonic regime of strong light-matter interaction, accompanied by experimental results demonstrating strong-coupling CPA behaviour.  The sample consists of a metallo-dielectric photonic crystal coupled to a multi-quantum-well heterostructure; the output intensity is normalized to the intensity of each input. $\omega_0$, resonance frequency; $\gamma_{nr}$, non-radiative decay rate of the cavity mode; $\gamma_m$, decay rate of the matter subsystem. Panel a is reprinted with permission from Ref.~\citenum{Roger2015}, panel b is reprinted with permission from Ref.~\citenum{Zanotto2014}.}
\label{quantum}
\end{figure*}

With the rapid advancement of quantum optics technology and its promising applications in telecommunications and quantum computing~\cite{Zbinden_2002, Ladd_2010}, it is important to investigate how CPAs function in the quantum regime, such as few-photon or entangled-photon states.

The possibility of achieving complete absorption of single photons has long been understood in the context of lossy beam splitters~\cite{Caves1982,Barnett1998, Jeffers}.  Consider a beam splitter with two input ports and two output ports, denoted $a$ and $b$.  For a single-photon state, such as
\begin{equation}
  |\varphi\rangle = \frac{1}{\sqrt{2}}\Big( |1_a,0_b\rangle + e^{i\varphi} |0_a,1_b\rangle \Big),
\end{equation}
 incident on the beam splitter, the state leaving the beam splitter is $\hat{S}|\varphi\rangle$, where $\hat{S}$ is unitary if the beam splitter is lossless.  If the beam splitter is lossy, $\hat{S}$ can have a zero eigenvalue; the corresponding single-photon input state is completely absorbed into the beam splitter itself, yielding no output state~\cite{Chong2013,Huang2014,Radi2015}.  The process is non-unitary because it describes the transfer of quantum state amplitudes into parts of the Hilbert space not included in the model, that is,  the 'unmonitored' degrees of freedom responsible for lossiness~\cite{Caves1982}.

An experimental demonstration of coherent perfect absorption for entangled single-photon states was realized~\cite{Roger2015},  starting with the preparation of the  single-photon state $|\varphi\rangle$ using spontaneous parametric down-conversion of a laser source together with a lossless 50/50 beam splitter (Fig.~5a).   The state $|\varphi\rangle$ was then directed onto a lossy metasurface with 50\% single-sided absorption (equivalent to a classical thin-film CPA).  The sample position was  modulated to vary the phase $\varphi$; the resulting nearly-perfect absorption was similar to that of a classical CPA (Fig.~5a).  In another recent experiment, coherent absorption was demonstrated for surface plasmon-polaritons in the quantum limit \cite{Science2017}, although non-entangled input states $\left| {{1_a},{1_b}} \right\rangle $ were used and perfect absorption was not achieved.

The situation is more complicated for multiple-photon states, such as a symmetric 'N00N state' of the form $2^{-1/2}\left( \left| N_a,0_b\right\rangle  + \left| 0_a,N_b \right\rangle\right)$, which is a superposition of $N$ photons in one interferometer arm and $N$ photons in the other.  For a two-photon N00N state, either both photons are completely absorbed by the beam splitter, or both photons pass through~\cite{Jeffers}.  This phenomenon has been observed using a multilayer graphene film as a lossy beam splitter~\cite{Roger2016}.

Another extension of the CPA concept into the quantum regime involves quantum nonlocality.  In a recent experiment \cite{SociArxiv}, a polarization-entangled photon pair was generated, and one photon directed into a two-arm interferometer containing a thin-film polarization-sensitive CPA.  Performing a polarization measurement on the external 'idle' photon collapsed the polarization state of the 'signal' photon inside the interferometer, allowing for nonlocal control over the photon absorption.

Several researchers have employed quantum models to describe physical limitations to coherent perfect absorption.  First, the performance of a CPA is limited by noise~\cite{Schoelkopf_2010}, which may be of thermal or quantum nature; even at zero temperature, there can be substantial quantum noise within the CPA's absorption bandwidth, originating from spontaneous emission induced by the incident coherent radiation~\cite{Chong2013}.  Second, the simple picture of a CPA as a linear absorber is accurate only if the absorbing medium is not saturable. If the absorbing medium is implemented as an ensemble of two-level systems, then absorption saturation may play a role.  Owing to the dispersive nature of a two-level medium, when gain is replaced by absorption, not only does the imaginary part of $\varepsilon(r)$ change sign, but so does the real part, whereas time-reversal symmetry requires the transformation $\varepsilon \to {\varepsilon ^*}$.  As a result, the CPA frequency is generally different from the lasing frequency in the equivalent system~\cite{Longhi2011}.  The inherent nonlinearity of a two-level medium also leads to bistability and hysteresis in the input-output characteristic of the system. Qualitatively similar behaviour was observed for a cavity quantum electrodynamic treatment of the problem~\cite{QED}.

Although the absorption processes giving rise to CPAs typically lie within the regime of weak light-matter interaction, CPAs can also be realized in strongly coupled systems, such as polaritonic cavities (Fig.~5b).  Using temporal coupled mode theory, it can be shown that coherent perfect absorption can be achieved by satisfying the strong critical coupling condition $\gamma_r=\gamma_{nr}+\gamma_{m}$, where $\gamma_r$ and $\gamma_{nr}$ are the radiative and non-radiative decay rates of the cavity mode, and $\gamma_m$ is the decay rate for the matter subsystem.  A strong-coupling CPA was  experimentally demonstrated~\cite{Zanotto2014} using a metallo-dielectric photonic crystal (patterned with gold stripes) coupled to a multi-quantum-well heterostructure with a resonant intersubband transition (Fig.~5b).

\section*{Applications}
\label{sec:Applications}

\begin{figure*}
\includegraphics[width=1\textwidth]{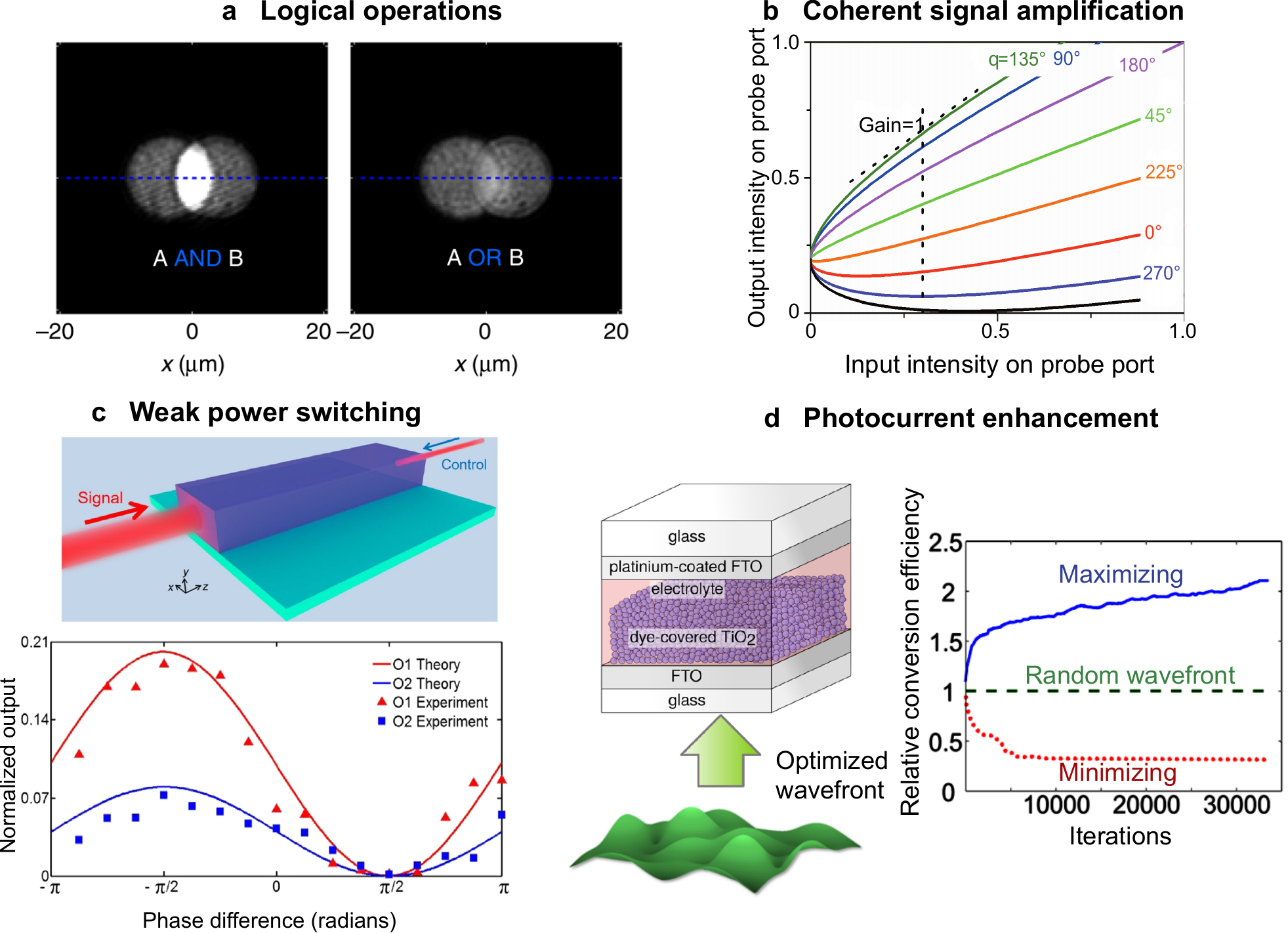}
\caption{\textbf {Applications of coherent perfect absorbers.} (a) Boolean operations can be implemented using a metasurface coherent perfect absorber (CPA).  Two coherent input beams form the beam spots, which do not overlap exactly; the intensity in the overlap region depends on the relative phase of the inputs, and serves as the output of the logic gate.    A similar scheme can be used for pattern recognition \cite{Zh-2017}.  (b) A two-port device can act as a coherent amplifier: with fixed input to the control port, the input-output relation at the probe port is nonlinear, and the effective gain depends on the relative phase of the inputs, $q$.   (c) A power-asymmetric CPA. The sketch shows a device illuminated by two beams of different intensities; the measured output power shows that a signal beam (O1) is modulated by a weaker ($\approx 30\%$ intensity) control beam O2.  The sample is a silicon-on-insulator waveguide near a $\mathcal{PT}$-breaking point.   (d) Photocurrent modulation through wavefront shaping.  The incident light passes through a spatial light modulator and enters a dye-sensitized solar cell containing strongly scattering nanoparticles.  Upon wavefront optimization by a genetic algorithm, the photocurrent can be increased by up to a factor 2 or decreased by up to a factor 0.3, as shown on the plot. FTO, fluorine-doped tin oxide.  Panel a is reprinted with permission from Ref.~\citenum{Zheludev16}, panel b is reprinted with permission from Ref.~\citenum{Fang2015}, panel c is reprinted with permission from Ref.~\citenum{Feng16}, panel d is reprinted with permission from Ref.~\citenum{Liew2016}.  }
\label{appl}
\end{figure*}

The phenomenon of coherent perfect absorption may be useful in a variety of applications, ranging from all-optical data processing to the exploitation of enhanced photovoltaic effects.  In the context of data processing, a CPA allows an optical signal to be strongly modulated by another coherent optical signal, without the need of a material nonlinearity.  Consider a two-port system satisfying the scattering relations  in Eq. (\ref{eq1}).  If both input amplitudes are scaled by a factor $\alpha$, the output fields are also scaled by $\alpha$.  However, if one input amplitude (a control beam) is fixed while the other (a probe beam) is varied, the response becomes essentially nonlinear. This suggests a route towards the realization of linear optical switches, modulators and logical gates~\cite{Fang2014, Fang2015, Zheludev16, LogicAPL}. CPAs have also been proposed for pulse restoration and coherence filtering of optical signals~\cite{Zhang2012}.

A prototype CPA-based device performing different boolean operations (AND, OR, XOR, and NOT) has been developed using a CPA setup to construct a desired truth table exploiting various phase shifts between the two incident beams~\cite{Fang2015,Zheludev16}. The realization of the AND and OR operations is illustrated in Fig.~6a.

By exploiting the fact that the output from the probe port varies nonlinearly with the control beam, it is possible to design a small-signal amplifier without incorporating nonlinear elements~\cite{Fang2015} (Fig.~6b).  The effective differential gain, $G=dI_{\rm out}/dI_{\rm in}$, varies as a function of the relative phase between the two input beams and can reach large values as the signal beam intensity approaches zero.

The idea of controlling light with light was further developed, and processing of binary images and pattern recognition were demonstrated \cite{Zh-2017}. Similarities and differences between binary images were detected by exploiting the coherent interaction of input test and target images, resulting in either transmisison or absorption of specific image elements. CPAs may also be employed for novel imaging schemes, such as volumetric imaging with reduced aberrations~\cite{PTlens}.

$\mathcal{PT}$-symmetric structures operating near the $\mathcal{PT}$-breaking point exhibit various interesting behaviours, such as strongly asymmetric reflection~\cite{Lin2011,Feng2012,Wu2016}.  This regime is associated with strongly asymmetric CPA eigenmodes, which allow the manipulation of a strong signal beam by a much weaker control beam (Fig.~6c). Such asymmetric switching has been experimentally demonstrated  in a periodically modulated Si waveguide, with observed extinction ratios of up to 60 dB for a 3:1 intensity ratio between signal and control beam~\cite{Feng16}.

Coherent absorption effects may prove very useful for improving photocurrent generation in photoelectrochemical systems~\cite{Liew2016}.  In such systems, the collection efficiency depends on the position at which charges are created.  Using the principle of coherent enhancement of absorption, incident waveforms can be tailored to enhance or suppress absorption.  This effect was demonstrated in a dye-sensitized solar cell, in which a six-fold modulation of the photocurrent was observed~\cite{Liew2016} (Fig.~6d).

Similarly, metasurface-based CPAs can be used to to control photoexcitation in photonic nanostructures.  An optical-frequency CPA was realized with a 2D periodic array of metallic nanopyramids on a silica substrate, and used to demonstrate coherent control over the photoluminescence intensity of a dye layer covering the nanoantennas~~\cite{Gomez16}.
Also, because coherent absorption experiments involve a delicate interplay between geometry, material absorption and incident waveform, they may provide a useful technique for non-destructive measurements of material absorptivity and other sample characteristics \cite{Graphene2017,Feng16}.

The field of acoustic metamaterials, which enable flexible control of sound waves, has been rapidly developing in recent years \cite{AcousticReview}. Following a few theoretical suggestions~\cite{Wei2014,Song2014}, an acoustic CPA was recently  experimentally demonstrated using decorated membrane resonators \cite{Null2017}.  Acoustic CPAs have a host of intriguing potential applications, such as noise control.

\section*{Discussion and Outlook}

We have discussed the fundamental principles of coherent perfect absorption, how CPAs can be implemented in various types of photonic structures and some examples of applications. CPAs have been realized in a wide variety of geometries, including slabs, gratings, metasurfaces, thin films, complex multi-port cavities and guided-mode structures compatible with integrated nanophotonics applications.  Different techniques have been devised for increasing or decreasing the absorption bandwidth and for achieving perfect absorption under strongly asymmetric inputs or non-planar incident waves.   CPAs have also been extended to the quantum optics regime.

Several of these advances were inspired by the concept of CPA as the time-reverse of a laser at threshold.  As a corollary to the principle of laser physics that ``anything will lase if you hit it hard enough'' \cite{SchawlowMemoirs}, practically any photonic structure can act as a CPA, once an appropriate amount of loss is introduced.  It may be interesting to look for CPA counterparts of various complex laser structures, such as photonic crystal slab lasers and random lasers.  However, unlike a laser, where the lasing mode emerges through self-organization \cite{Haken}, perfect absorption requires the supply of exactly the right incident waveform. It is unclear whether there can be any useful CPA analog of laser self-organization --- it could be perhaps found in the presence of nonlinear absorptive media, as suggested by several theoretical studies~\cite{Longhi2011,QED,Reddy2013a,Reddy2013b,Reddy2014}.  Furthermore, interesting variants of the time-reversal concept have been formulated for nonlinear processes such as wave mixing and parametric oscillations \cite{longhi2011a,Schackert2013,Zheng2013}.
For instance, the idea of a coherent perfect absorber for waves of different frequencies, acting as the time-reverse of an optical parametric oscillator, was put forward~\cite{longhi2011a}.

As we have seen, promising applications for coherent perfect absorption are beginning to be explored.  Looking ahead, we expect CPAs or CPA-like devices to be useful is situations where coherent optical signals need to be compactly and efficiently processed.  Consider, for instance, the prospect of a complicated photonic circuit that employs coherent light for linear low-power optical signal processing or computing.  If only non-absorptive interferometric effects are used, whenever an unwanted signal product has to be removed from the system --- for example an optical bit that plays no further role in a computation --- an unmonitored output port must be inserted.  This may complicate the circuit design if the unwanted signal occurs deep within the circuit.  Integrated CPA devices could resolve this difficulty by serving as efficient 'sinks' for optical signals \cite{Bruck,Park2015,Rothenberg2016, Intracavity}.  In the quantum optics limit, the ability to control and remotely trigger perfect absorption of single- or few-photon states could likewise be useful for quantum information processing \cite{SociArxiv}.

Another broad class of applications for CPAs involves the exploitation of the sharp intensity contrast between CPA and non-CPA eigenmodes.  The most extreme examples are CPA-laser devices \cite{Longhi2010, Chong2011, Gu2016,Wong2016}, in which the output intensity is zero for the CPA eigenmode and very high for other inputs (limited only by gain saturation).  CPA-lasers have been demonstrated in simple two-port configurations; in the future, such devices could allow for sensitive interferometric measurements, particularly of low-intensity optical signals.


\begin{acknowledgments}
D.G.B.~and T.S.~acknowledge support from the Knut and Alice Wallenberg Foundation. D.G.B.~acknowledges support from the Russian Foundation for Basic Research (project no.~16-32-00444).  T.S.~acknowledges financial support from Swedish Research Council (Vetenskapsomr\r{a}det, grant no.~2012-0414).  A.A.~and A.K.~acknowledge support from the Air Force Office of Scientific Research with grant no.~FA9550-17-1-0002 and the Welch Foundation with grant no.~F-1802. C.Y.D.~is grateful to A.~D.~Stone, L.~Ge, and A.~Cerjan for numerous stimulating and deep discussions, and acknowledges support from the Singapore MOE Academic Research Fund Tier 2 grant no.~MOE2015-T2-2-008, and the Singapore MOE Academic Research Fund Tier 3 grant no.~MOE2011-T3-1-005.
\end{acknowledgments}

\bibliography{CPA_bib}

\end{document}